\documentclass[letter]{ieice}

\usepackage[dvips]{graphicx}
\newcommand{\oo}{\bot}            
\newcommand{\pp}{\top}            

\newcommand{\pst}{\mbox{\raisebox{-0.01cm}{\scriptsize $\wedge$}\hspace{-4pt}\raisebox{0.16cm}{\tiny $\mid$}\hspace{2pt}}}
\newcommand{\gneg}{\neg}                  
\newcommand{\mli}{\rightarrow}                     
\newcommand{\cla}{\mbox{\large $\forall$}}      
\newcommand{\cle}{\mbox{\large $\exists$}}        
\newcommand{\mld}{\vee}    
\newcommand{\mlc}{\wedge}  
\newcommand{\ade}{\mbox{\Large $\sqcup$}}      
\newcommand{\ada}{\mbox{\Large $\sqcap$}}      
\newcommand{\add}{\sqcup}                      
\newcommand{\adc}{\sqcap}                      

\newcommand{\tlg}{\bot}               
\newcommand{\twg}{\top}               

\newtheorem{theoremm}{Theorem}[section]
\newtheorem{conditionss}{Condition}[section]

\newtheorem{definitionn}[theoremm]{Definition}
\newtheorem{lemmaa}[theoremm]{Lemma}
\newtheorem{notationn}[theoremm]{Notation}
\newtheorem{propositionn}[theoremm]{Proposition}
\newtheorem{conventionn}[theoremm]{Convention}
\newtheorem{examplee}[theoremm]{Example}
\newtheorem{remarkk}[theoremm]{Remark}
\newtheorem{factt}[theoremm]{Fact}
\newtheorem{exercisee}[theoremm]{Exercise}
\newtheorem{questionn}[theoremm]{Open Problem}
\newtheorem{conjecturee}[theoremm]{Conjecture}

\newenvironment{lemma}{\begin{lemmaa}}{\end{lemmaa}}

\newcounter{itemno}

\newcounter{itemno1}

\newcounter{itemno2}

\newcounter{exno}

\newcounter{defno}

\newcommand{\oprove}{\vdash\kern-.6em\lower.7ex\hbox{$\scriptstyle O$}\,}

\newcommand{\pderivation}{{\cal P}\kern -.1em\hbox{\rm -derivation}}
\newcommand{\pderivationl}{{\cal P}\kern -.1em\hbox{\em -derivation}}
\newcommand{\pderivable}{{\cal P}\kern -.1em\hbox{\rm -derivable}}
\newcommand{\pderivablel}{{\cal P}\kern -.1em\hbox{\em -derivable}}
\newcommand{\pderivations}{{\cal P}\kern -.1em\hbox{\rm -derivations}}
\newcommand{\pderivability}{{\cal P}\kern -.1em\hbox{\rm -derivability}}

\newsavebox{\lpartfig}
\newsavebox{\rpartfig}

\newenvironment{exmple}{
 \begingroup \begin{tabbing} \hspace{2em}\= \hspace{3em}\= \hspace{3em}\=
\hspace{3em}\= \hspace{3em}\= \hspace{3em}\= \kill}{
 \end{tabbing}\endgroup}

\newcommand{\colw}{CoLweb}
\field{}
\title{Logical Pseudocode: Connecting Algorithms with Proofs}
\authorlist{
\authorentry{Keehang Kwon}{m}{labelA}
\authorentry{Hyung Joon Kwon}{m}{labelB}
}
\affiliate[labelA]{  
 DongA University. khkwon@dau.ac.kr}

\affiliate[labelA]{KAIST.
hyungjoon65@kaist.ac.kr}

\received{2003}{1}{1}
\revised{2003}{1}{1}
\finalreceived{2003}{1}{1}

\begin{document}
\maketitle
\begin{summary}
Proofs (sequent calculus, natural deduction) and imperative algorithms (pseudocodes) 
are two well-known coexisting concepts.  Then what is their relationship?
Our answer is that \\

   \[ imperative\ algorithms\ =\  proofs\ with\ cuts \]

 \noindent This observation leads to a generalization to  pseudocodes which we call 
  {\it logical pseudocodes}. It is similar to 
  natural deduction proof of computability logic\cite{Jap03,Jap08}. 
  Each statement in it corresponds to a proof step in natural deduction.
  Therefore, the merit over pseudocode is that
 each statement is guaranteed to be correct and  safe with respect to the 
  initial specifications.   It can also be seen as an extension to  
  computability logic web (\colw)  with forward reasoning
  capability.
\end{summary}


\section{Introduction}\label{sec:intro}

Pseudocode (and  imperative languages) is  low-level and
 unregulated. For example, the assignment statement
$x\ =\ E$ is not regulated at all and is considered  harmful.
 As a consequence, it is difficult to tell whether a pseudocode is
correct or safe.

Of course, it is possible to use high-level logic  languages. 
These languages make code much safer and much easier to
verify
 but often this requires some sacrifice
in the performance. This is mainly because they are not able to utilize 
lemmas. 

What are the remaining alternatives then? We believe
\colw \cite{Kee2020} is a best choice for expressing algorithms.
 It is a  promising model of multi-agent programming where
 agents exchange services. It can be viewed as a   
high-level, bottom-up version of natural deduction proof of computability logic.
In contrast to logic languages, the main feature of \colw\
is a full utilization of lemmas. In other words, the main idea of 
\colw\  is to create high-level, regulated, safe and lemma-based
codes.
However, while implementing \colw\  in a distributed setting poses little problems,
implementing it in a non distributed setting is somewhat complicated.

For this reason, we introduce {\it logical pseudocode} -- as a companion to
\colw\  -- which can be
seen as a version of CoLweb with top-down, centralized control.
Logical pseudocode is much easier to implement than CoLweb in
a non-distributed environment. It can be seen as  a major evolution of pseudocode.

\section{Problems with Pseudocode}

Let us consider an array-based implementation of the fibonacci sequence.
In the traditional imperative approach,  it  can be  specified as: \\

fib(n) =  \%  an array-based implementation\\

\begin{exmple}
\>  a[0] = 1; \\
\>  a[1] = 1; \\
\>  for i = 0 to n-2 \\
 \>\>  a[i+2] = a[i+1]+a[i] \\
 \>  return a[n];
  \end{exmple}
  
  In the above, $a[0] = 1$, etc  is called an assignment statement.
 
 The above code has at least two weaknesses:
  
  \begin{itemize}
  
  \item It is not clear whether the code is correct, and
  
  \item It is not clear whether the code is safe.

  \end{itemize}
  
  That is, the machine always executes pseudocodes, regardless of their correctness and safety.
  This is a serious problem which  leads to various researches such as  
   proof-carrying code. As a different direction of this trend, we introduce logical assignment statements 
   as a replacement of
  assignment statements.
  
  \section{Preliminaries}\label{s2}

In this section a  brief overview of CoL is given. 

There are two players: the machine $\pp$ and the environment $\oo$.

There are two sorts of atoms: {\em elementary} atoms $p$, $q$, \ldots to represent elementary games, and {\em general atoms} $P$, $Q$, \ldots to represent any, not-necessarily-elementary, games. 

\begin{description}
\item[Constant elementary games]  $\twg$ is always a true proposition, and $\tlg$ is always a false proposition.

\item[Negation]
 $\gneg$ is a role-switch operation: For example, $\gneg (0=1)$ is true,
while $(0=1)$ is false.

\item[Choice operations]
The choice group of operations:  $\adc$, $\add$, $\cla$ and $\cle$ are defined below.

$\ada xA(x)$ is the game where, in the initial position, only $\oo$ has a legal move which consists in 
choosing a value for $x$. After $\oo$ makes a move $c\in\{0,1,\ldots\}$, 
the game continues as $A(c)$. 
$\cla xA(x)$ is similar, 
only here the value of $x$ is invisible. $\ade$ and $\cle$ are 
symmetric to $\ada$ and $\cla$, with
the  difference that now it is $\pp$ who makes an initial move.

\item[Parallel operations]
Playing $A_1\mlc\ldots\mlc A_n$ means playing the $n$ games concurrently.  In order to win,  $\pp$ needs to win in each of $n$ games. Playing  $A_1\mld\ldots\mld A_n$ also means playing the $n$ games concurrently.  In order to win,  $\pp$ needs to win  one of the games. To indicate that a given move is made in the $i$th component, the player should prefix it with the string ``$i.$".  
The operations $\pst A$ means an infinite parallel
game $A\mlc\ldots\mlc A\mlc\ldots$.
 To indicate that a given move is made in the $i (i>1)$th component, we assume the player should 
 first replicate $A$ and then prefix it with the string ``$i.$".

\item[Reduction]
 $A\mli B$ is defined  by $\gneg A\mld B$.
Intuitively, $A\mli B$ is the problem of reducing $B$ ({\em consequent}) to $A$ ({\em antecedent}).  

\end{description}

\section{Introducing Directories }\label{sec:intro}

Logical formulas  are inadequate for locating subformulas. 
Our approach to achieving this effect is through the use of 
{\it directories}. For example, consider the following  directory
definition.

\[ /m = p(a) \]

\noindent where $/m$ is a directory name and $p(a)$ is a formula.
In this case, we call $p(a)$ its ``content''.
Alternatively, we can view $/m$ as an agent and $p(a)$ as its
knowledgebase. A directory is a sequence of subdirectories and formulas. For example,

\[ /m = (/m_1 = \ldots, \ldots,/m_n = \ldots, F_1\ldots, F_n) \]

\noindent where $/m$ is a directory, each $/m_i$ is a subdirectory and each $F_i$ is a file.
We also introduce a {\it class directory/agent} which is a combination of the form

\begin{itemize}

\item $\cla x\ /m(x) = F$, or

\item $ \pst\ /m = F.$

\end{itemize}

Our directory system is very flexible and is designed
to represent both formulas and cirquents.
For example, $/n = !/m \land !/m$ represents that the directory $/n$
contains $p(a) \land p(a)$.
 Here $!/m$ is intended to read as ``a copy of the
 content of $/m$. In contrast, $/o = /m \land /m$ represents that
 $/o$ contains a cirquent
 $p(a) \land p(a)$ where two $p(a)$s in $/o$ and $p(a)$ in $/m$
 are $shared$.


As another example, consider the following  recursive directory
definition.

\[ /m(0) = q \]
\[ /m(s(X)) = p \land !/m(X) \]

\noindent Given this definition, $p \land (p \land (p \land q)))$ can be represented simply as $ /m(s(s(s(0))))$.  We assume in the above that
$s$ is the number-successor function.

Thus, we propose the notion of {\it directorized}
formulas.  They are  formulas enhanced with 
directories. These formulas are better-suited to structuring large
formulas such as pigeonhole
principle formulas.
It is interesting to note that directories also play the role of global
variables
in
imperative languages and much more.

 \section{Logical Assignment Statements}

 In the sequel, we focus on a single machine with local memories. 
 Note that a single machine with local memories can be viewed as a multiagent
 system where each agent shares a single CPU.
 
 To overcome the problems mentioned in the previous section,
 we introduce  {\it logical assignment statements}  of the form 
 
 \[  /x\ =\ F^{l_1,\ldots,l_n} \]
 
\noindent     where $/x$ is an agent/location, $F$ is a logical formula and each $l_i$ is an agent. We call $F^{l_1,\ldots,l_n}$ {\it querying knowledgebase}. We call the above statement {\it well-formed} if
$F$ is a logical consequence of knowledgebases at $l_1,\ldots,l_n$.
 
 Agents operate in three modes: idle, reactive and proactive. Agents in reactive mode
 means that they are busy processing service requests from other agents.
 Agents in proactive mode
 means that they invoke querying knowledgebase to  other agents.

 {\it Executing} $/x$ means that $/x$ is in proactive mode.
 That is, the machine tries to solve the query $F$ using knowledgebases at $l_1,\ldots,l_n$. If it succeeds, it binds $/x$ to $F'$ to which $F$ evolves. Otherwise, it reports failure.

 For example, consider the following:
 
\begin{exmple}
 $/x\ =\ p(0,1)$ \\
 $/y\ =\  \ade wp(0,w)^{ /x }$ \\
$/z\ =\  p(0,5)^{ /x }$ \\
 \end{exmple}

\noindent In the above, note that $/y$ is well-formed but $/z$ is not.
Now executing $/y$ means the machine tries to solve the goal using $p(0,1)$.
This will succeed and  bind $/y$ to $p(0,1)$. On the contrary,  executing $/z$ will not be allowed.
Note that destructive logical assignments are disallowed.

\section{Forward  Construct}

In pseudocode, for example, $x=4;y=x$ imposes the following sequence  between 
two statements:  execute $x=4$  and then execute $y=x$.
This {\it forward} ordering supports forward reasoning in theorem proving.
Note that forward chaining is often simple to implement, while implementing
backward chaining requires a chain of stacks. 

From this viewpoint, we add a similar construct of the form

\[ /x\ =\ F; /y\ =\ G. \]
\noindent The above means the following: to execute $/y$, execute $/x$ first.

 For example, consider the following:

\begin{exmple}
$/x\ =\ p(0,1)$ \\
$/y\ =\ \ade y p(0,y)^{/x}$; \\
$/z\ =\ \ade z p(0,z)^{/y}$ \\
\end{exmple}
\noindent Executing $/z$ requires to execute $/y$ first. After execution, we obtain
the following:

\begin{exmple}
$/x\ =\ p(0,1)$ \\
$/y\ =\  p(0,1)$ \\
$/z\ =\  p(0,1)$ \\
\end{exmple}

The forward  relation can be generalized to the forward quantifier of the following form:

\[ for\ i_m^n\  /x[i]\ =\ F(i) \]
\noindent which means

\[ /x[m]\ =\ F(m); \ldots ;\ /x[n]\ =\ F(n). \]

 To activate $/x[k]$, the machine does the following: \\
 
 (step 1) It generates the following:

\begin{exmple}
\>\>$/x[m]\ =\ F(m)$; \\
\>\>$\vdots$\\
\>\>$/x[k]\ =\  F(k)$; \\
\>\>$ for\ i_{k\!+\!1}^n\  /x[i]\ =\ F(i)$\\
\end{exmple}
\noindent 

(step 2) The machine executes $/x[m],\ldots,/x[k]$ in that order.

Finally, a well-formed {\it logical pseudocode}  $P$ is a set of well-formed logical assignment statements.
Given $P$, execution tries to solve a query using a mix of backward reasoning and  forward reasoning.

\section{Adding Induction}

Let $R$ be a set of initial assumptions or axioms.
 To deal with a query $/q = (\ada xF(x))^R$, we need to introduce
mathematical induction to logical pseudocode. 
There are several reasonable systems for dealing with induction.
One simple example is the constructive induction \cite{Jap08} is of the 
form

\[ F(1) \mlc \ada x(F(x)\mli F(x+1)) \mli \ada xF(x). \]

By assuming that $F(X)$ is stored at a location $/a[X]$,
the above can be rewritten in a more economical form, which we call $IND$.  That is, we assume that
the programmer generates the following code:

\begin{exmple}
 $/a[1] =  F(1)^R$\\
  $/istep =  for\ i_{2}^\infty\  /a[i] =  F(i)^{/a[i-1],R}$  \\
  $/q = (\ada xF(x))^{IND,/a[1],/istep}$ \\
\end{exmple}
\noindent In the above, $/istep$ denotes an agent which handles the inductive steps.

The above IND is called a special induction which is a restricted case of general induction.
The general induction GIND can be similarly defined but is not shown here for
simplicity. It, however,  will be used in the sequel.

As an example,  consider $/fib = \ada x\ade yfib(x,y)^R$.  Then it  can be
written as: \\

\begin{exmple}
 $/a[1] =  \ade yfib(1,y)^R$\\
$/a[2] = \ade yfib(2,y)^R$ \\
 $/istep =  for\ i_3^\infty\  /a[i] =  \ade y fib(i,y)^{/a[i-1],/a[i-2],R}$  \\
$/fib = (\ada x\ade wfib(x,w))^{GIND,/a[1],a[2],/istep}$ \\
\end{exmple}
\noindent In the above, $R$ is a set of rules for the fibonacci sequence.

Note that the above code  is well-formed. Now let us execute the query $/fib$. 
This expression waits for an input. Suppose the $/fib$ types in 4.
The machine then  invokes
 $\ade y fib(4,w)$ to the agent $/a[4]$.
Due to the for-loop, the machine creates  and
executes two agents $/a[3], /a[4]$
in that order. This in turn invokes executing $/a[1],/a[2]$. Now the above code becomes the following:

\begin{exmple}
 $/r[1] = fib(1,1)$ \\
 $/r[2] = fib(2,1)$\\
$/r[3] = \cla x,y,z (fib(x,y)\mlc fib(x\!+\!1,z)) \mli fib(x\!+\!2,y\!+\!z)) $ \\
  $/a[1] =  fib(1,1)$ \\
$/a[2] =fib(2,1)$ \\
$/a[3] = fib(3,2)$ \\
$  /a[4] = fib(4,3)$ \\
 $ for\ i_5^\infty\  /a[i] =  \ade y fib(i,y)^{/a[i\!-\!1],/a[i\!-\!2],R}$ \\
$/fib = \ade wfib(4,w))^{/a[4]}$ \\
\end{exmple}
\noindent Now the agent $/fib$ stores $fib(4,3)$ as desired and execution terminates.

\section{An Example}

As an example, we consider the fibonacci sequence.
We  assume that the user invokes queries to the agent $/fib$.
  Then $fib$ can be
written as: \\

\begin{exmple}
 $/r[1] = fib(1,1)$ \\
 $/r[2] = fib(2,1)$\\
$/r[3] = \cla x,y,z (fib(x,y)\mlc fib(x\!+\!1,z)) \mli fib(x\!+\!2,y\!+\!z)) $ \\
 $/a[1] =  \ade yfib(1,y)^{/r[1]}$\\
$/a[2] = \ade yfib(2,y)^{/r[2]}$ \\
 $/istep =  for\ i_3^\infty\  /a[i] =  \ade y fib(i,y)^{/a[i-1],/a[i-2],/r[3]}$  \\ 
$/fib = (\ada n \ade yfib(n,y))^{GIND,/a[1],/a[2],/istep}$ \\
$/query = (\ada x\ade wfib(x,w))^{/fib}$ \\
\end{exmple}

Note that the above code is well-formed. In particular, $/query$ is well-formed, as it is a logical consequence of $/fib$.
Now let us execute $/query$. 
This expression waits for an input. Suppose $/query$ types in 4 for $x$.
The machine then tries to solve it  relative to the program in $/fib$. Then $/fib$ must select 4 for $n$.
The rest proceeds as explained in the preceding section.

\noindent 
 The above code is very concise but has
some interesting features:

\begin{enumerate}
  
\item It supports automatic memoization.
  
\item It supports a mix of backward reasoning and forward reasoning (via for-loop).
\end{enumerate}

\section{Thinking Theorem Proving in a Bigger Paradigm}\label{ttp}

A sequence of well-chosen lemmas plays a key role in classical theorem proving such as
natural deduction (ND), sequent calculus with cuts (LK+cut) and Coq.
Yet, finding lemmas can be extremely challenging. For this reason, 
effective strategies and tactics for finding lemmas 
have been studied but with little progress.

One way to approach this problem is to embed theorem proving in a bigger paradigm.
For example,   logical pseudocode (and \colw) -- based on
computability logic -- provides some useful
solutions in this regard. That is, it allows us to specify 
{\it a set of lemma candidates} instead of a single lemma. This makes life simpler.
To be specific, it extends lemmas to include the following:

\begin{itemize}

\item $(lemma_1 \add \ldots \add lemma_n )^{l_1,\ldots,l_k}$: 
In this case, the machine tries to find $lemma_i$ which is a logical consequence of knowledgebases at $l_1,\ldots,l_k$.

\item $\ade x lemma(x)^{l_1,\ldots,l_k}$: In this case, the machine tries to find a value $c$ of $x$ such that
$lemma(c/x)$  is a logical consequence of knowledgebases at $l_1,\ldots,l_k$

\end{itemize}
\noindent In the above $lemma$ is a first-order formula.
For example,  an infinite set of lemmas such as
$\ade x fib(100,x)$ is now allowed.

In conclusion, logical pseudocode (and \colw) can be seen as a nondeterministic
version of ND and a better alternative to ND.  We can apply this idea
to the cut formula of LK or to the tactics of Coq to 
obtain a nondeterministic version of LK+cut or a nondeterministic version of 
tactics.

\section{Conclusion}

Our ultimate goal is to implement the computability
logic web  which is a promising approach to
reaching general AI.  New ideas in this paper -- forward reasoning
construct  -- will be useful for practical implementations.

\bibliographystyle{ieicetr}


\end{document}